\documentclass[prb,twocolumn,showpacs,amsmath,amssymb,superscriptaddress]{revtex4-1}
\usepackage{amsmath}
\usepackage{graphicx, nicefrac}
\usepackage{float}
\usepackage{siunitx}

\usepackage{dcolumn}
\usepackage{bm,graphicx}
\usepackage{etoolbox}

\usepackage[colorlinks]{hyperref}

\usepackage[dvipsnames,usenames]{color}
\usepackage[version=3]{mhchem} 

\usepackage{breakurl}
\usepackage{color}
\usepackage{amsmath}
\usepackage{xcolor}
\usepackage{soul}

\newcommand{\mt}{\ce{MoTe2}}
\newcommand{\icm}[1]{\SI{#1}{\per\centi\meter}}
\newcommand{\K}[1]{\SI{#1}{\kelvin}}
\newcommand{\mev}[1]{\SI{#1}{\milli\electronvolt}}
\newcommand{\ev}[1]{\SI{#1}{\electronvolt}}

\newcommand{\h}{\hbar}

\newcommand{\e}{\varepsilon}
\newcommand{\s}{\sigma}

\newcommand{\om}{\omega}

\newcommand{\D}{\Delta}

\begin{document}

\bibliographystyle{apsrev}

\title{Low-energy excitations in type-II Weyl semimetal T$_d$-MoTe$_{2}$ \\
evidenced through optical conductivity}

\author{D. Santos-Cottin}
\email[]{david.santos@unifr.ch} 
\affiliation{Department of Physics, University of Fribourg, 1700 Fribourg, Switzerland}

\author{E. Martino}
\affiliation{Department of Physics, University of Fribourg, 1700 Fribourg, Switzerland}
\affiliation{IPHYS, EPFL, CH-1015 Lausanne, Switzerland}

\author{F. Le Mardel\'e}
\affiliation{Department of Physics, University of Fribourg, 1700 Fribourg, Switzerland}

\author{C. Witteveen}
\affiliation{Department of Chemistry, University of Z\"urich, CH-8057 Z\"urich, Switzerland}
\affiliation{Physik-Institut der Universitat Z\"urich, CH-8057 Z\"urich, Switzerland}

\author{F. O. von Rohr}
\affiliation{Department of Chemistry, University of Z\"urich, CH-8057 Z\"urich, Switzerland}
\affiliation{Physik-Institut der Universitat Z\"urich, CH-8057 Z\"urich, Switzerland}

\author{C.~C.~Homes}
\affiliation{Condensed Matter Physics and Materials Science Department, Brookhaven National Laboratory, Upton,
   New York 11973, USA}

\author{Z.~Rukelj }
\affiliation{Department of Physics, University of Fribourg, 1700 Fribourg, Switzerland}
\affiliation{Department of Physics, Faculty of Science, University of Zagreb, Bijeni\v{c}ka 32, HR-10000 Zagreb, Croatia}

\author{Ana  Akrap}
\email[]{ana.akrap@unifr.ch} 
\affiliation{Department of Physics, University of Fribourg, 1700 Fribourg, Switzerland}

\date{\today}
\begin{abstract}
Molybdenum ditelluride, \mt, is a versatile material 
where the topological phase can be readily tuned by manipulating the associated structural phase transition.
The fine details of the band structure of \mt, key to understanding its topological properties, have proven difficult to disentangle experientially due to the multi-band character of the material.
Through experimental optical conductivity spectra, we detect two strong low-energy interband transitions. Both are linked to excitations between spin-orbit split bands. The lowest interband transition shows a strong thermal shift, pointing to a chemical potential that dramatically decreases with temperature.
With the help of {\it{ab initio}} calculations and a simple two-band model, we give qualitative and quantitative explanation of the main features in the temperature-dependent optical spectra up to $400$~meV.
\end{abstract}
\maketitle

%
Molybdenum ditelluride, \mt, belongs to the rich and diverse family of transition metal dichalcogenides (TMDs). Both in bulk and few-layer form, TMDs are intensely studied for many of their interesting properties: excitons, superconductivity, band-gap tuning by thickness, as well as for their possible applications in electronics, optoelectronics, spintronics and valleytronics \cite{arora_valley_2016, yin_ultrahigh_2016, lin_ambipolar_2014, pradhan_field-effect_2014,keum_bandgap_2015}.

The semimetallic phases of group IV (Mo, W) TMDs can crystallize in the monoclinic 1T$'$ and orthorhombic T$_d$ structures. Those materials have attracted a lot of attention due to their predicted topological properties such as the quantum spin Hall effect, or presence of Weyl fermions,\cite{sun_prediction_2015, wang_mote_2016, chang_prediction_2016, soluyanov_type-ii_2015} which can be tuned by switching from the T$'$ to the distorted T$_d$ phase by temperature, strain or light pulses.\cite{Zhang2019}
Most recently, it was shown that the superconductivity becomes strongly enhanced as \mt\ is taken to its monolayer limit. The superconducting transition sets in at 8 K, sixty times higher than in the bulk compound, where $T_c = 0.13$ K.\cite{Rhodes2019,qi_superconductivity_2016}
Similarly to T$_d$-WTe$_2$, T$_d$-€™\mt\ is predicted to be a type-II Weyl semimetal with a strong spin-orbit coupling arising from inversion symmetry breaking. Four pairs of Weyl nodes are expected in the band structure, at \mev{6} and \mev{59} above E$_F$,\cite{sun_prediction_2015} on top of tilted conically dispersing bands.
The electronic properties of this phase have been addressed by band structure calculations, angle-resolved photoemission spectroscopy (ARPES), quantum oscillations and magneto-transport measurements. 
A large and non saturating magneto-resistance may be understood in terms of a quasi-perfect compensation of charge carriers at low temperature,\cite{zhou_hall_2016, rhodes_bulk_2017, pei_mobility_2018} similar to T$_d$-WTe$_2$.\cite{Homes2015}
Fermi arcs have indeed been observed by ARPES, with different surface band dispersions corresponding to different Weyl nodes \cite{sakano_observation_2017,Tamai2016}.
However, it has proven difficult to probe the low-energy band structure directly by experiments. Understanding the detailed band structure is also particularly important for the observed superconductivity enhancement in monolayer \mt. 

In this paper, we address the low-energy band structure of T$_d$-\mt\, by means of detailed infrared spectroscopy measured down to \mev{2}, in conjunction with optical response functions calculated from the band structure.
We identify the low-energy valence band structure by comparing specific features of the optical spectroscopy measurements with the electron bands calculated by density functional theory (DFT), and the optical conductivity calculated from an effective low-energy model. The unique sensitivity to both intraband (Drude-like) and interband transitions allows us to disentangle the details of the band structure in the very low, mili-electronvolt energy range. The temperature dependence of the optical response shows an important renormalization of the spectral weight up to \ev{1} in function of temperature. A strong broadening of the Drude term with the increase in temperature accompanies the emergence of a peculiar low-energy interband transition, with a pronounced thermal shift. This suggests that the chemical potential strongly depends on temperature.
\\
%
%
Millimeter-sized high-quality single-crystals of 1T$'$-\mt\ were synthesized using a self flux method.\cite{Guguchia_2017}
Electrical resistivity was measured in a Physical Property Measurement System from Quantum Design as a function of temperature. The sample was measured using a four-probe technique in a bar configuration in the $ab$-plane.\\
The optical reflectivity was determined at a near-normal angle of incidence with light polarized in the $ab$-plane for photon energies ranging between \mev{2} and \ev{1.5} (16 and \icm{12000}), at temperatures from 10 to \K{300}. 
The single crystal was mounted on the cold finger of a He flow cryostat and absolute reflectivity was determined using the \textit{in-situ} coating technique \cite{Homes_1993}.
The data was complemented by an ellipsometry measurement up to \ev{6.3} (\icm{51000})  at room temperature.
The complex optical conductivity was obtained using a Kramers-Kronig transformation from the reflectivity measurements. At low frequencies, we used a Hagen-Rubens extrapolation. For the high frequencies, we completed the reflectivity data using the calculated atomic X-ray scattering cross sections \cite{Tanner_2015} from 10 to \ev{60} followed by a $1/\omega^{4}$ dependence. 
\\
The electronic properties of MoTe$_2$ in the orthorhombic $Pmn2_1$ (31) phase have been calculated using density functional theory (DFT) with the generalized gradient 
approximation (GGA) using the full-potential linearized augmented plane-wave (FP-LAPW) method \cite{Singh} with local-orbital extensions \cite{Singh91} in the WIEN2k implementation \cite{Wien2k}.  
The unit cell parameters have been adjusted and the total energy calculated both with and without spin-orbit coupling; while spin-orbit coupling lowers the total energy, it does not significantly affect the structural refinement. Once the unit cell has been optimized, the atomic fractional coordinates are then relaxed with respect to the total force (spin-orbit coupling is not considered in this step), typically resulting in residual forces of less than 0.2 mRy/a.u. per atom.  This procedure is repeated until no further improvement is obtained.  The electronic band structure has been calculated from the optimized geometry with GGA and spin-orbit coupling.
%
%

\begin{figure}[htb]
	\includegraphics[width=8.5cm]{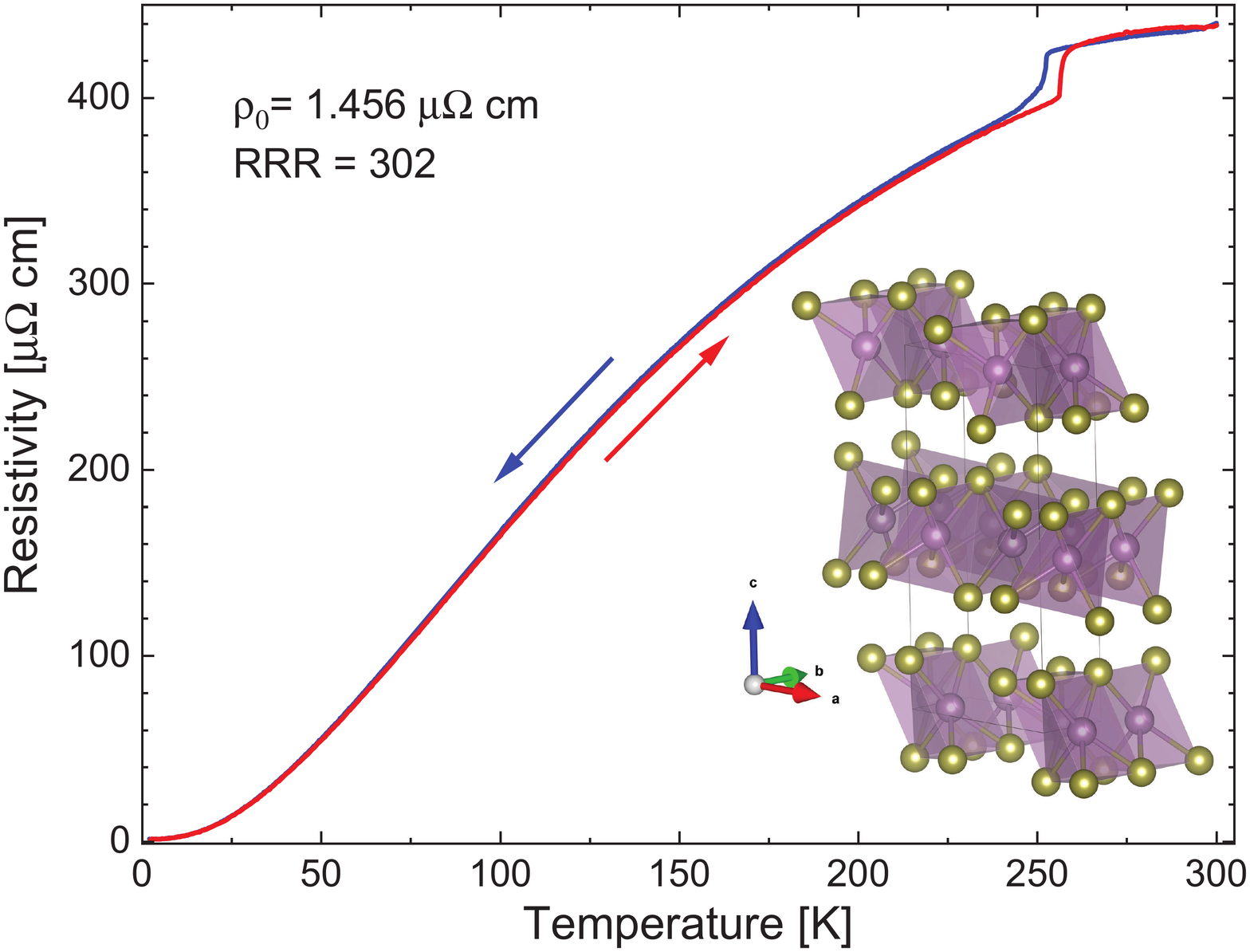}
  	\caption{Temperature dependence of the $a$-axis resistivity of \mt\ is shown for cooling (blue) and warming up (red). The inset shows the lattice structure of \mt, where yellow spheres represent tellurium atoms and violet spheres molybdenum atoms.}
  	\label{fig1}
\end{figure}

Figure \ref{fig1} shows the temperature-dependent electrical resistivity of \mt, with current applied along the $a$ axis. Resistivity was measured in cooling and heating the sample, shown in blue and red respectively. The resistivity is typical of a semimetallic system, characterized by a strong decrease as the temperature is reduced. The very large residual resistivity ratio RRR $=\rho_{\K{300}} / \rho_{\K{2}}\simeq 300$, with $\rho_{\K{2}} = 1.46\cdot10^{-6}$ $\Omega$~cm, indicates the high quality of our single crystal, with values very similar to the recently investigated WTe$_2$.\cite{Homes2015}
The abrupt change of the resistivity slope at \K{250} is due to a phase transition between the high-temperature monoclinic 1T$'$ phase ($P2_1/m$ space group) and the low-temperature orthorhombic T$_d$ phase ($Pmn2_1$ space group). This phase transition has been investigated through different techniques, mainly  X-ray diffraction\cite{{clarke_XRD_1978},{dawson_electronic_1987},{kim_origins_2017}} and transport measurements.\cite{{yan_investigation_2017},{hughes_electrical_1978},{qi_superconductivity_2016}} Only recently have the experiments confirmed that the low temperature T$_d$ phase breaks inversion symmetry, leading to a Weyl semimetal phase.\cite{{zhang_raman_2016}, {berger_temperature-driven_2018}}

The inset to Fig.~\ref{fig1} shows the room-temperature 1T$'$-phase crystal structure of \mt.\cite{brown_crystal_1966} Tellurium atoms, in yellow, form distorted octahedra which surround the molybdenum atoms. The octahedral distortion is due to an $ab$-plane displacement of the metal ion, which moves to the center of the octahedra in the low-temperature T$_d$ phase.
Both the 1T$'$ and T$_d$ phase of \mt\ are layered, quasi two-dimensional structures. Each layer is a sandwich of three atomic sheets, Te-Mo-Te, arranged in a covalently bonded 2D-hexagonal configuration. Layers are connected to each other through weak van der Waals coupling.\cite{{clarke_XRD_1978},{dawson_electronic_1987}}

Below $\sim$ \K{50}, the resistivity follows a quadratic dependence in temperature, $\rho =\rho_0 + AT^2$, with $A = 2.18\cdot 10^{-2}\ \mu\Omega$ cm~K$^{-2} $, similar to a previous report. \cite{zandt_quadratic_2007} In a large number of Fermi liquids, the prefactor $A$ is directly related to the Fermi energy, falling onto a universal curve.\cite{lin_scalable_2015} 
This phenomenological extension of Kadowaki-Woods relation points to a fairly low Fermi energy in \mt, estimated to $\sim 15$~meV. 

\begin{figure}[htb]
	\includegraphics[width=8.5cm]{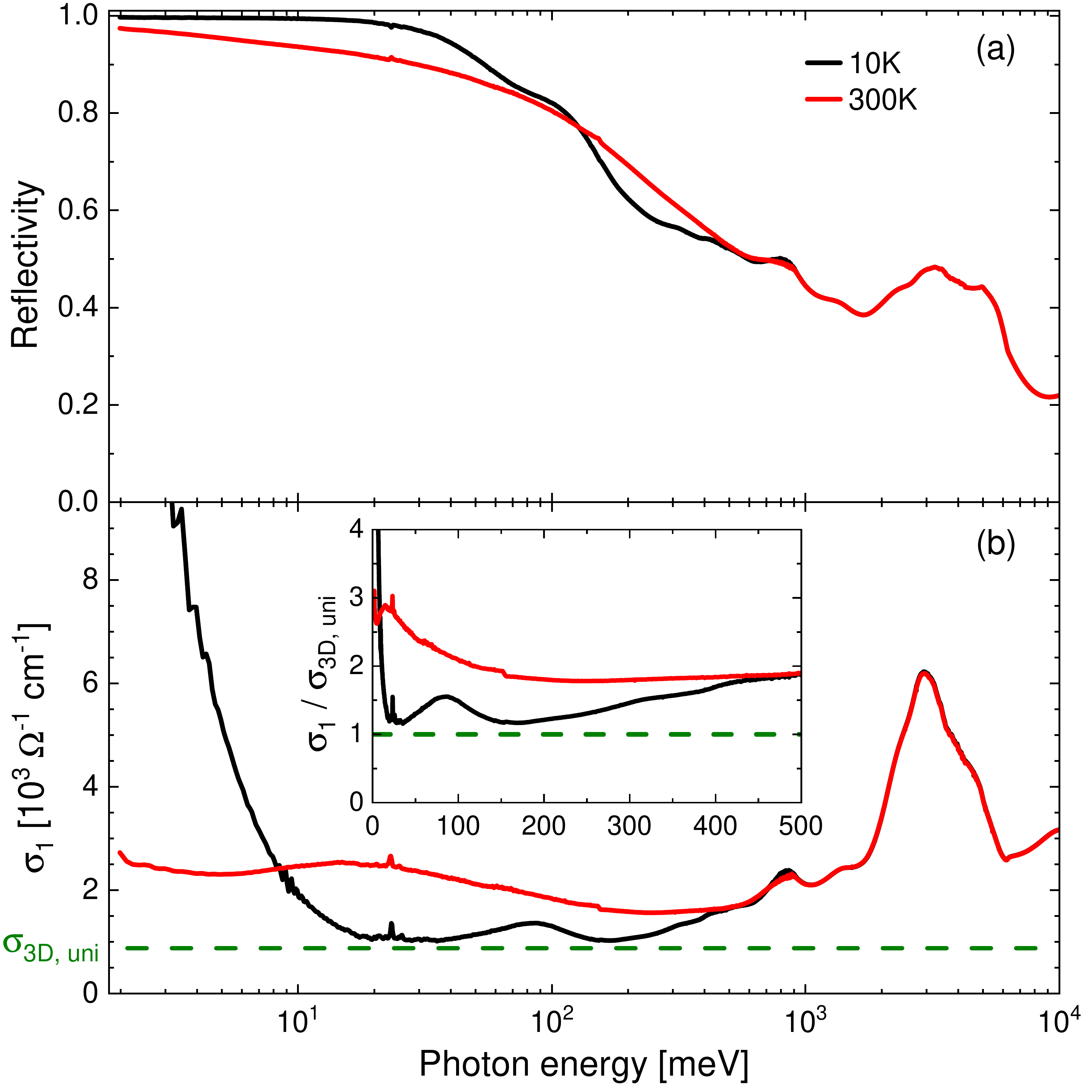}
  	\caption{(a) The in-plane reflectivity in the full spectral range is shown for  $T=$~\K{300} and \K{10}, in red and black, respectively. (b) The real part of the optical conductivity $\sigma_1(\om)$, is shown in the same photon energy range. The horizontal dashed line represents the 3D universal conductance, $\sigma_{3D, \mathrm{uni}}$, as described in the main text. 
	Inset shows the ratio $\sigma_1/\sigma_{3D, \mathrm{uni}}$ below \mev{500} on a linear photon energy scale.}
  	\label{fig2}
\end{figure}

Figure \ref{fig2} shows (a) the reflectivity $R$ and (b) the real part of optical conductivity, $\sigma_1(\om)$, at \K{10} and \K{300}, for a broad range of photon energies. The reflectivity behaves as expected in a semimetal,  with $R(\om) \rightarrow 1$ in the low-energy limit, $\omega \rightarrow 0$. 
At \K{300}, the low energy reflectivity increases continuously, faster than linear with the decrease of energy. In contrast, at \K{10} the reflectivity shows a saturation plateau approaching $R \sim1 $ for photon energies below \mev{20}. This plateau translates into a much higher conductivity than at \K{300}, which agrees with the transport data. No temperature dependence of reflectivity can be discerned for photon energies above \ev{0.5}. 

At low energies and low temperature, $\sigma_1(\omega)$ exhibits a very narrow Drude contribution superimposed on a flat electronic background. A much broader Drude component is observed at \K{300}, giving rise to a very weakly frequency-dependent $\sigma_1(\omega)$. The Drude scattering rates are low, $\hbar/\tau \sim 1$~meV at 10 K, and $\sim 5$~meV at room temperature. A large change happens in the Drude plasma frequency, which drops by a factor of 2.6 from \K{10} to room temperature, leading to an almost sevenfold decrease in the Drude weight.
Such a dramatic loss of Drude contribution from \K{10} to \K{300} leads to a strong spectral weight transfer from far infrared to mid infrared, evident in Fig.~\ref{fig2}b. The drop in the Drude strenght is fully consistent with a very large drop in resistivity with cooling. If \mt\, is treated as a multiband system, then a fit with two Drude components is more meaningful. This fit results in a narrow Drude component superimposed on a broad one. In this approach, at 10~K the Drude scattering rate of the narrow component is $1.5$~meV, and $247$~meV for the broad component. The Drude plasma frequencies are 780~meV and 1240~meV respectively, and this is consistent with a nearly compensated system.

Similarly to the reflectivity measurements, above \ev{0.5} we observe no significant temperature dependence of $\sigma_1(\om)$. At around \ev{3}, there is a strong peak corresponding to a high energy interband transition, possibly a transition along the SX direction in the Brillouin zone, which points between the Te--Te layers.
At high energies our data overall agrees with a recent optical study\cite{Kimura2019}. However, our ability to reach much lower photon energies with a better experimental resolution give us access to the critical energy range needed to address the previously unseen features in the low energy band structure. 

Due to its low symmetry crystal structure, T$_d$-\mt\ has many Raman-active phonon modes, 17 modes are experimentally observed.\cite{Ma_Ramansym_2016, zhang_raman_2016}  Absence of inversion symmetry dictates that all these phonon modes also be infrared-active.
However, a simple empirical force-field model indicates that only two of these modes have a significant dipole moment. As a result, in $\sigma_1(\omega)$ there is only one clear infrared-active phonon mode, appearing at \mev{23.4} (\icm{188.5}) for \K{10}. This mode softens slightly as temperature rises, and is seen at \mev{23.1} (\icm{186.5}) for \K{300}. 
From Raman spectra, a phonon mode of likely $B_1$ symmetry is expected around \mev{24}. 

Much more prominent in the $\sigma_1(\omega)$ spectra are several distinct, low-lying interband transitions. The narrow Drude contribution sits on top of a strong background of interband transitions.
In a layered system such as \mt, generally one expects a weak interlayer dispersion. It is then interesting to compare $\sigma_1(\omega)$ in the interband region (above \mev{10}) to the dynamical universal sheet conductance, which can be determined from the relation $\sigma_{3D,\mathrm{uni}} = G_0/d_c = e^2/(4\hbar d_c)$. Here, $G_0$ is the conductance quantum, and $d_c$ the interlayer distance.\cite{Kuzmenko_UniConductance_2008} 
In Fig.~\ref{fig2}b, the dashed line shows the three-dimensional (3D) universal sheet conductance given the interlayer Mo-Mo distance of $d_c=c/2 =6.932\ \mathrm \AA$, where $c$ is the lattice parameter at low temperatures. The value $\sigma_{3D,\mathrm{uni}} \sim 1000\ \Omega^{-1}$cm$^{-1}$ appears to be in reasonable agreement with the low-temperature $\sigma_1(\om)$ for photon energies between 10 and \mev{500}. This may imply that in a first approximation, an in-plane Dirac-like band dispersion in \mt\ is responsible for most of the observed interband transitions, while the interlayer dispersion remains very weak.

Two well-defined peaks at finite energies are observed in $\sigma_1(\omega)$ shown in Fig.~\ref{fig2}b. These peaks are both linked to low-energy interband transitions. One of them is centered around \mev{90} at \K{10}, while another, broader interband transition can be seen at \K{300} at \mev{20}. To better understand the origin of these two interband transitions, it is important to look at their detailed temperature dependence. 

\begin{figure}[htb]
  		\includegraphics[width=8cm]{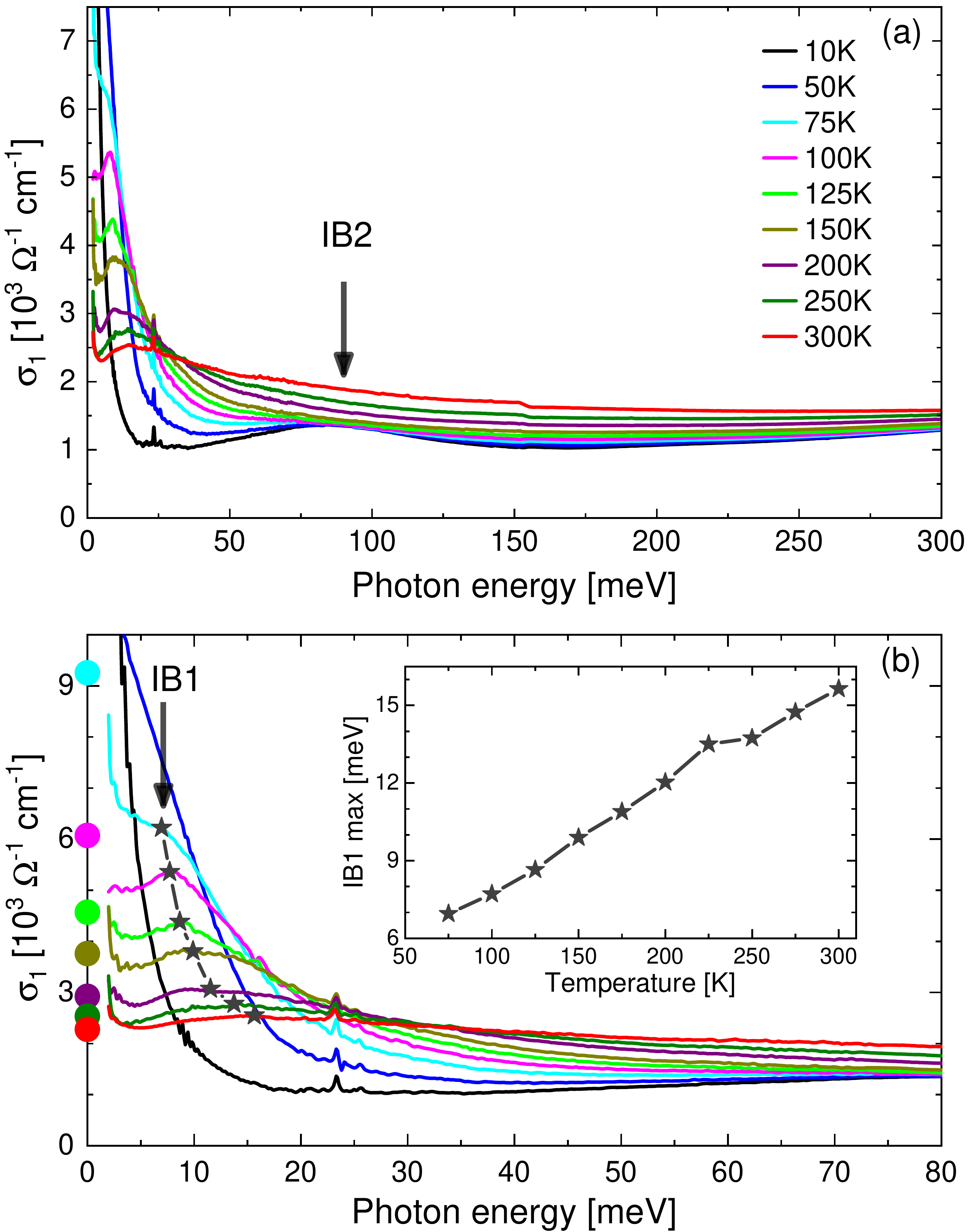}
  		\caption{(a) The real part of the optical conductivity, $\sigma_1(\om)$. 
		(b) Optical conductivity in the very far-infrared region, focusing on the lower-energy interband transition. $\sigma_{dc}$ values are extracted from the resistivity measurement in Fig.~\ref{fig1} at various temperatures and are represented by large colored circles. Interband transition IB1 is marked by stars. The inset in (b) shows the energy of the peak of IB1 as a function of temperature.}
  	\label{fig3}
\end{figure}

Figure \ref{fig3}a shows the detailed temperature dependence of the real part of the optical conductivity $\sigma_1(\om)$ in the midinfrared energy range up to \mev{300}. 
The temperature dependence of $\sigma_1(\om)$ clearly shows a steady narrowing of the Drude contribution as temperature decreases, consistent with a gradual loss of carriers and their reduced scattering time. Excellent agreement between the low energy  $\sigma_1(\om)$ and the $\sigma_{dc}$ values, extracted from data in Fig.~\ref{fig1}, confirms the low-energy behavior of the optical conductivity. 

Overlapping with the Drude contribution, we can unequivocally isolate a narrow and strongly temperature-dependent peak which we call IB1 (Fig.~\ref{fig3}b). Due to its shape, its finite energy, and its temperature dependence, this peak in $\sigma_1(\om)$ can only be attributed to an interband transition.
The peak position shifts from \mev{7} at \K{75}, to \mev{16} at \K{300}  (see inset in Fig.~\ref{fig3}b), while its intensity diminishes with increasing temperature.
There seems to be a subtle change in the temperature behavior of the IB1 peak around \K{250}, the temperature where the structure changes from the high-temperature 1T$'$ phase to the low-temperature T$_d$ phase. Between \K{75} and \K{200}, the temperature dependence of the IB1 maximum appears to be linear or possibly parabolic. 
A second, broader interband peak is visible at \mev{90} at \K{10} (Fig.~\ref{fig3}a), and we refer to it as IB2. In contrast to the strongly blue-shifting low-energy peak IB1, the position of the higher peak IB2 seems to very slightly red-shift as the temperature increases. The temperature-induced broadening of the Drude component effectively washes out this higher interband transition, rendering it indistinguishable above \K{100}. 

\begin{figure*}[htb]
  		\includegraphics[width=14cm]{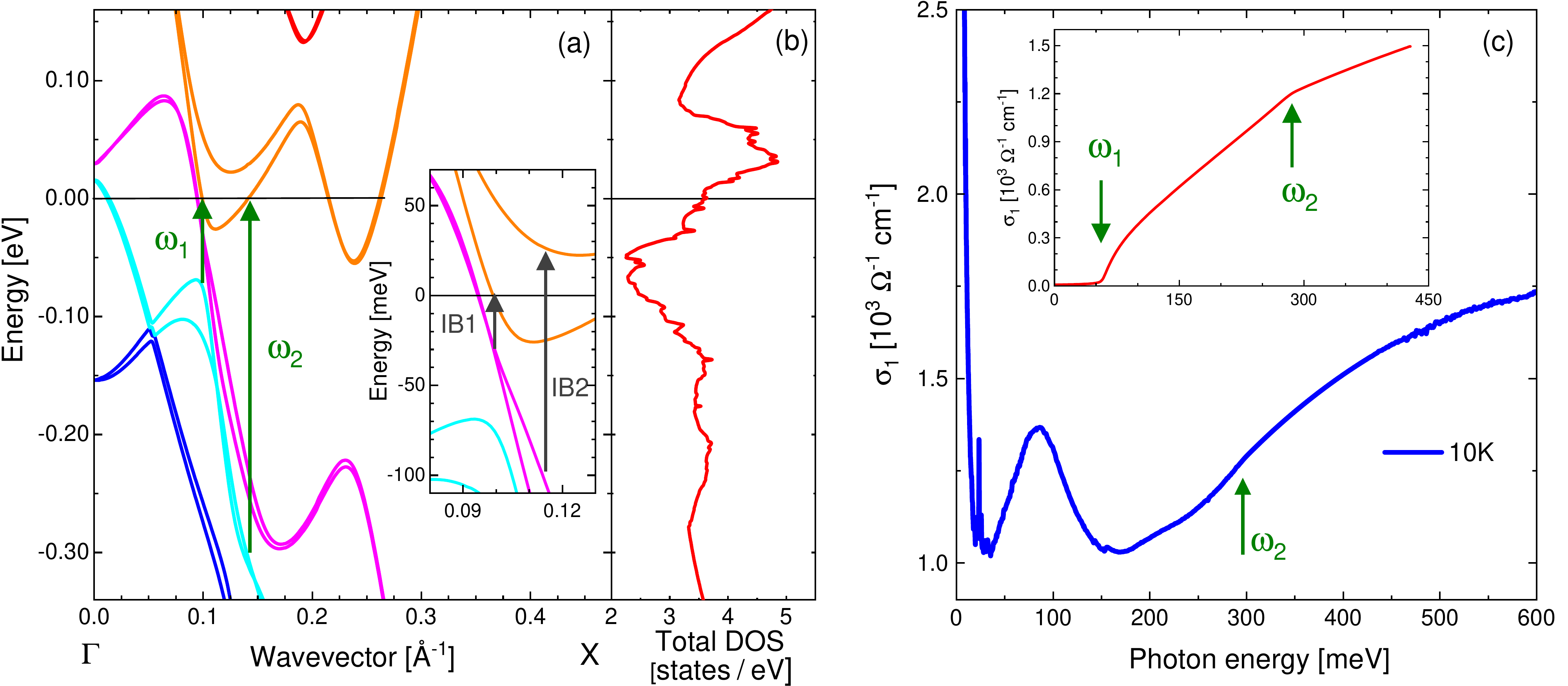}
  		\caption{(a) DFT calculation of band structure for the orthorhombic phase of \mt. Inset: enlarged low-energy region. (b) Total density of states (DOS) around the Fermi level.
		(c) Experimental $\sigma_1(\om)$ at 10 K. Inset:  the theoretically calculated interband contribution, limited to the tilted quasilinear bands.}
  	\label{fig4}
\end{figure*}

Interband contribution to the optical conductivity is linked to the band structure through its dependence on the joint density of states (JDOS). Very roughly, $\sigma_1 \propto$ JDOS$(\om)/\om$.
This relation means that we can identify the possible origins of IB1 and IB2 by comparing our optical measurements to the band structure of \mt, and thereby clarify the details of its low-energy band structure.
To this purpose, Fig.~\ref{fig4}a shows the DFT calculation of the low-energy band structure of orthorhombic \mt. 
It reaffirms that the material is a multiband conductor.\cite{Crepaldi2017,sakano_observation_2017}
From the band structure in the $\Gamma-X$ direction, we can identify that IB1 must be a transition between levels that are in the vicinity of Weyl points.

Similarly, for IB2, judging by the low temperature dependence, this peak may be attributed to the transitions between the steeply dispersing (magenta) band, and the upper parabolic (orange) band. This assignment is consistent with a $\sim 100$~meV energy separation between the bottom of the upper parabolic band and the steep (magenta) band; this energy difference corresponds to the maximum JDOS.

It is rather unusual for an interband transition to show such a strong thermal shift as what we see for IB1. The strong shift cannot be caused by a change in the band structure, as it is not expected to change below 250~K.
The most reasonable way to explain the thermal shift of IB1 is to allow that the chemical potential $\mu(T)$ moves very strongly as a function of temperature. Generally, when increasing $T$, $\mu(T)$ will shift to the energy where the density of states is lower, so as to preserve the charge neutrality. In our case, this means $\mu(T)$ should shift downwards as the temperature increases, since DOS is monotonically decreasing at the Fermi level (Fig.~\ref{fig4}b).
IB1 shifts by \mev{10} from 75~K to 300~K, which corresponds to $\sim \Delta T/2$. If this shift is caused by a chemical potential change, in other words by a temperature-dependent Pauli blocking, one would expect the shift to behave like $\propto T^2$. This is consistent with our data below \K{200}, see inset of Fig.~\ref{fig3}b.

Because the band structure is complex, it is impossible to exactly determine the partial contributions to the total interband $\sigma_1(\omega)$ from the transitions IB1 and IB2.
These interband transitions are given by intricate sums in reciprocal space.\cite{Martino2019} Despite this limitation, we believe the assignment in Fig.~\ref{fig4}a is justifiable.
Generally, for any interband transition we expect to have a higher JDOS and hence a stronger optical transition when the two involved bands are nearly parallel; in the limiting case, this is a van Hove singularity. 

Above the IB2 peak, there are additional features in the optical spectra that imply a specific band character. At the energy $\omega_2=290$~meV there is a kink, followed by nearly square root energy dependence, $\sigma_1 \propto \sqrt{\omega}$. Such a kink is characteristic of the optical response of a tilted 3D Dirac system. In contrast, in a 3D Dirac system the optical conductivity at $\omega > \omega_2$ has a linear dependence, $\sigma_1 \propto \omega$.
As seen in Fig.~\ref{fig4}a, the DFT shows that the Fermi level crosses the upper of the the two gapped tilted quasilinear bands. 
The interband transition between these bands lead to a kink in $\sigma_1(\omega)$ at $\omega_2$, as well as a $\sqrt{\omega}$ dependence of $\sigma_1(\omega)$.
To show this explicitly, we construct an effective  $2\times2$ Hamiltonian assuming a free-electron-like behavior in the $z$ direction and a linear energy dependence in $xy$ ($ab$) plane:
 \begin{equation} \label{ham1}
 \hat{H}_0 = \h wk_x\s_0+  \h v k_x\s_x + \h v k_y\s_x + (\D+\xi(z))\s_z.
\end{equation}
Here, $\s_{x,y,z}$ are Pauli matrices, $\s_0$ is the unity matrix, $w$ is the tilt parameter, $v$ is the velocity in the $x$ and $y$ direction, and $2\D$ is the energy band gap.
For the out-of-plane direction we assume $\xi(z)=\hbar^2k_z^2/2m^* $, where $m^*$ is   the effective mass. 
This choice is made based on weakly dispersing bands in the $z$ direction, which implies $m^* \gg m_e$.
Interband $\sigma_1(\om,T)$ can be numerically evaluated from Eq.~\ref{ham1}, using the well-known form of the conductivity tensor.\cite{Martino2019} The result is shown in the inset of Fig.~\ref{fig4}c.
A signature of the tilted conical (quasi linear) bands may be identified in the two subtle kinks at $\omega_1$ and $\omega_2$ in $\sigma_1(\om)$, indicated by arrows. If Fermi energy measured from the bandgap
middle is $\e_F > \Delta$, we have
 a way to determine the upper Pauli blocking energy, $\hbar\om_{2} \approx 2\e_F/(1+w/v)$.
DFT gives the bandgap $2\Delta = 40$~meV, the Fermi level (measured from the middle of the band gap) $\e_F=45$~meV, the
tilt $w = -4.8\times 10^5$~m/s and  the velocity $v = 6.7\times 10^5$~m/s.
For $\om > \om_2$, the optical conductivity is described by\cite{Martino2019}
\begin{equation}\label{op5}
 {\rm{Re}} \, \s^{vc}_{xx}(\omega \geq \om_2,T=0) = \frac{\s_0}{\pi} \frac{\sqrt{m^*}}{\hbar}\sqrt{\h \omega - 2\D},
\end{equation}
where $\sigma_0 = e^2/(4\hbar)$. Comparison with experimental $\sigma_1(\omega)$ gives the effective mass $m^* = 13 m_e$, which justifies the flat band assumption.

%
%
In conclusion, through a combined use of detailed infrared spectroscopy and effective modelling, we show that  the low energy dynamical conductivity in \mt\ is dominated by complex interband transitions, due to a rich band structure at the Fermi level. 
The intraband (Drude) contribution to conductivity is greatly dependent on temperature.
We observe a narrow low-energy interband transition, whose pronounced temperature-dependence points to a strong temperature dependence of the chemical potential in \mt.
The tilted quasilinear bands, and an associated quickly dispersing band, are responsible for much of the low-energy interband transitions. We detect a subtle signature of the tilted conical dispersion.

%
%
We would like to thank C. Bernhard for the use of experimental setup, M. M\"uller, and A.B. Kuzmenko for their comments and suggestions, and N. Miller for kind help.
A.~A. acknowledges funding from the  Swiss National Science Foundation through project PP00P2\_170544. Z.R. was funded by the Postdoctoral Fellowship of the University of Fribourg. F.O.v.R. was funded by the  Swiss National Science Foundation through project PZ00P2\_174015.
Work at BNL was supported by the U.S. Department of Energy, Office of Basic Energy Sciences, Division of Materials Sciences and Engineering under Contract No. DE-SC0012704.
%
%
\bibliography{biblio}

\end{document}